\documentclass[aps,prb,twocolumn,amsmath,amssymb,showpacs,groupedaddress]{revtex4}

\usepackage{bm}
\usepackage{graphicx}

\begin{document}


\title{Hexagonal Warping Effects on the Surface Transport in Topological Insulators}

\author{C. M. Wang}
\email[]{cmwangsjtu@gmail.com}
\author{F. J. Yu}
\affiliation{School of Physics and Electrical Engineering, Anyang
Normal University, Anyang 455000, China}

\date{\today}

\begin{abstract}
We investigate the charge conductivity and current-induced spin
polarization on the surface state of a three-dimensional topological
insulator by including the hexagonal warping effect of Fermi surface
both in classical and quantum diffusion regimes. We present general
expressions of conductivity and spin polarization, which are reduced
to simple forms for usual scattering potential. Due to the hexagonal
warping, the conductivity and spin polarization show an additional
quadratic carrier density dependence both for Boltzmann contribution
and quantum correction. In the presence of warping term, the surface
states still reveal weak anti-localization. Moreover, the dielectric
function in the random phase approximation is also explored, and we
find that it may be momentum-angle-dependent.
\end{abstract}


\maketitle

\section{introduction}
Topological insulator (TI) has been attracting a great deal of
research both experimentally and theoretically in the past few
years\cite{Kane2005,qi2008,qi2010topological} due to its potential
applications in topological quantum computation\cite{Fu2008} and
spintronics.\cite{Garate2010} A TI has a full energy gap in bulk,
while there are gapless surface states stable against weak disorder
and weak interaction unless time-reversal symmetry is
broken.\cite{Wu2006helical,xu2006stability} In particular, the
number of its Dirac points is odd according to a no-go
theorem,\cite{Wu2006helical} in vivid contrast to the
graphene.\cite{novoselov2005two} Hence, many theoretical studies
focus on a class of TI where the surface states only consist of one
Dirac cone.

Experimentally, the surface state has been well confirmed by
angle-resolved photoemission spectroscopy
(ARPES).\cite{hsieh2008topological,Hsieh13022009,Chen10072009}
However, the transport observation of surface state in classical
diffusion regime meets an obstacle due to the large bulk-conduction
background. Recently, Checkelsky {\it et al.} claimed to have
isolated the surface band contribution in Bi$_2$Se$_3$ by
electrostatic gate control of chemical
potential.\cite{checkelsky2010surface} Nevertheless, this work is
questionable for the reason that the chemical potential is still in
the bulk conduction band.\cite{Dimitrie} Theoretically, D. Culcer
{\it et al.} investigated the two-dimensional surface charge
transport and obtained results analogous to the ones of
graphene.\cite{Dimitrie} Recently, Kim {\it et al.} showed that the
surface of thin Bi$_2$Se$_3$ were electrostatically coupled,
strongly.\cite{Kim} They observed the surface transport by using a
gate electrode to remove bulk charge carriers, completely, and well
demonstrated the theoretical prediction.\cite{Dimitrie} It is very
likely that this experiment has overcome the above obstacle. Also,
the surface anomalous Hall conductivity in TI was calculated using
quantum Liouville equation.\cite{culcer2010anomalous} The
electron-phonon scattering limited conductivity was investigated for
the surface state of a strong TI.\cite{giraud2011electron} Moreover,
the transverse magnetic heat transport was explored on the
topological surface.\cite{yokoyama2011transverse}

On the other hand, recently, the quantum corrections to charge
conductivity in topological surface states were also extensively
studied.\cite{checkelsky2010surface,Chen2010,HongTao2011,Minhao2011,lu2011competing,tkachov2011weak}
The surface states in quantum diffusion regime ($l_\phi\gg l$)
reveal positive weak localization correction, i.e. weak
anti-localization, which is related to the $\pi$ Berry
phase.\cite{lu2011competing} Here, $l$ and $l_\phi$ are the elastic
scattering length and the phase coherence length, respectively. In
contrast to graphene, where the weak anti-localization is suppressed
by the intervalley scattering,\cite{McCann2006} the surface states
of TI forbid this scattering process due to single Dirac cone and
many observations have confirmed this enhancement to electronic
conductivity.\cite{checkelsky2010surface,Chen2010,HongTao2011,Minhao2011}
Lu {\it et al.} found that the surface state of a TI shows a
competing effect of weak localization and weak anti-localization in
quantum transport due to magnetic doping.\cite{lu2011competing}

We have noted that, in most of these theoretical studies, only $\bm
k$-linear term in the spin-orbit interaction is present in the
effective Hamiltonian. However, in some TIs, for example
Bi$_2$Te$_3$, it was found that the shape of Fermi surface changes
from a circle to a hexagon, and then to a snowflake-like shape with
increasing the Fermi energy by ARPES\cite{Chen10072009} and scanning
tunneling microscopy\cite{Zhanybek2010} measurement. With the help
of a hexagonal warping term, this kind of band structure was
explained by Fu.\cite{Fu266801} Note that the states near the Fermi
surface are responsible for the transport properties at low
temperature. Hence, it is expected that the warping effect will
naturally play significant roles on surface transport when the Fermi
energy is high enough. So far, only one theoretical work involves
the warping effect on the weak anti-localization in the
literature.\cite{tkachov2011weak} Further, they simply replaced the
warping term by its angle average one, hence, the anisotropy of
energy spectrum due to warping is neglected, completely. Then they
acquired the same form of correction as the usual two-dimensional
electron gas with spin-orbital interaction. Consequently, it is
highly desirable to carefully study the warping effect on the
classical contribution and quantum correction to the surface
transport.

In this paper, we study the hexagonal warping effect on the charge
conductivity and current-induced spin polarization (CISP) in the
surface state of a three-dimensional TI. Considering nonmagnetic and
magnetic elastic carrier-impurity scattering, we discuss this
problem both in classical and quantum diffusive regimes. We also
investigate the warping effect on the dielectric function in the
random phase approximation (RPA). The structure of the paper is as
follows. In Sec. II, the effective Hamiltonian of the surface state
is given. By using a kinetic equation approach, the conductivity and
CISP in the classical transport regime in the presence of
nonmagnetic and magnetic scattering are calculated in Sec. III and
IV. In Sec. V, we discuss the quantum correction to conductivity and
CISP. A brief summary is given in Sec. VI.

\section{system and Hamiltonian}
By assuming particle-hole symmetry, the effective Hamiltonian of the
surface state of a TI including the hexagonal
warping effect has the following form:
\begin{equation}\label{ham}
\hat H_0=v_{\rm
F}(k_x\hat\sigma_y-k_y\hat\sigma_x)+\frac{\lambda}{2}(k_+^3+k_-^3)\hat\sigma_z.
\end{equation}
Here $v_{\rm F}$ and $\lambda$ are the Fermi velocity and the
hexagonal warping constant, $\hat \sigma_i$ ($i=x,y,z$) are the
Pauli matrices, and $k_\pm=k_x\pm ik_y$. A quadratic term $k^2/(2m)$
is in principle also present in the Hamiltonian with $m$ denoting
the effective mass of particle, but it is smaller than the $\bm
k$-linear and cubic terms due to the relation $2mv_{\rm
F}\gg\sqrt{v_{\rm F}/\lambda}$ in Bi$_2$Te$_3$. Consequently, in the
regime of density, $10^{13}\,{\rm cm}^{-2}<N<10^{14}\,{\rm
cm}^{-2}$, both the linear and cubic terms contribute significantly
to the transport quantities, and the quadratic term can be safely
neglected. Near the regime $\bm k=0$, the cubic correction is also
negligible. However, at high density, the cubic term makes the
energy spectrum of surface state angle-dependent and the Fermi
surface becomes snowflake-like.

The eigenenergies of the considered system \eqref{ham} can be found
$\varepsilon_{\bm k\mu}=(-1)^\mu\epsilon_{\bm k}$, where
\begin{equation}\label{}
\epsilon_{\bm k}=\sqrt{(v_{\rm F}k)^2+(\lambda k^3\cos3\theta_{\bm
k})^2}
\end{equation}
with the azimuthal angle of $\bm k$, $\theta_{\bm
k}=\tan^{-1}(k_y/k_x)$, and the index $\mu=1,2$. By introducing the angle
\begin{equation}\label{}
\beta_{\bm k}=\tan^{-1}\sqrt{\frac{\epsilon_{\bm k}-\lambda
k^3\cos3\theta_{\bm k}}{\epsilon_{\bm k}+\lambda k^3\cos3\theta_{\bm
k}}},
\end{equation}
the corresponding eigenstates $\varphi_{\bm k\mu}$ are written as
\begin{equation}\label{}
\varphi_{\bm k1}=\begin{pmatrix}
                   \sin\beta_{\bm k} \\
                   -i\cos\beta_{\bm k} e^{i\theta_{\bm k}} \\
                 \end{pmatrix},
\end{equation}
\begin{equation}\label{}
\varphi_{\bm k2}=\begin{pmatrix}
                   \cos\beta_{\bm k} \\
                   i\sin\beta_{\bm k} e^{i\theta_{\bm k}} \\
                 \end{pmatrix}.
\end{equation}
It should be noted that the above Hamiltonian \eqref{ham} can be
diagonalized into $H_0=U_{\bm k}^\dag \hat H_0 U_{\bm k}={\rm
diag}(\varepsilon_{\bm k1},\varepsilon_{\bm k2})$ with the help of
the local unitary transformation $U_{\bm k}=(\varphi_{\bm
k1},\varphi_{\bm k2})$. This transformation projects the system from
the spin basis to the eigenbasis of $\hat H_0$.

\section{classical transport in the presence of nonmagnetic scattering}
\subsection{kinetic equations}
In order to study the transport property of the surface state in the
classical diffusive regime, we limit our system to a spacial homogeneous one. First, we
consider the nonmagnetic carrier-impurity elastic scattering and focus
 on the charge transport at the Fermi level inside the bulk gap of TI. The
kinetic equation for the single particle distribution function in
the eigenbasis of $\hat H_0$, $\rho(\bm k)$, are constructed using
the nonequilibrium Green's function and is given by\cite{Haug}
\begin{equation}\label{}
\left(\frac{\partial}{\partial T}-e\bm E\cdot\nabla_{\bm
k}\right)\rho+e\bm E\cdot[\rho,U_{\bm k}^\dag\nabla_{\bm k}U_{\bm
k}]+i[\hat H_0,\rho]=-I_{\rm sc}.
\end{equation}
Here $\bm E$ is the electric field. It should be noted that here the
$\rho(\bm k)$ is a $2\times2$ matrix. In the lowest order of
gradient expansion, the scattering integral $I_{\rm sc}$ can be
written as $ I_{\rm
sc}=\int^T_{-\infty}dt'\big[\Sigma^rG^<+\Sigma^<G^a
-G^r\Sigma^<-G^<\Sigma^a\big](T,t')(t',T)$, with $\Sigma^{<,r,a}$
being the lesser, retarded and advanced self-energies in the
self-consistent Born approximation. We consider the scattering by
impurities at random positions $\{\bm R_\alpha\}$ has the form:
$\widetilde V(\bm r)=\sum_{\{\bm R_\alpha\}}V(\bm r-\bm R_\alpha)$.
Therefore, after impurity averaging,\cite{Haug} the self-energies in
eigenbasis of $\hat H_0$ reads $\Sigma^{<,r,a}(\bm k)=n_i\sum_{\bm
q}|V(\bm k-\bm q)|^2U_{\bm k}^\dag U_{\bm q} G^{<,r,a}(\bm q) U_{\bm
q}^\dag U_{\bm k}$, with $n_i$ denoting the impurity density and
$V(\bm q)$ being Fourier transform of $V(\bm r)$.

Further, we take the generalized Kadanoff-Baym ansatz\cite{Haug} and
ignore the collisional broadening to simplify the scattering
integral. Throughout this paper, we focus on the situation where the
Fermi energy $\varepsilon_{\rm F}$ is positive, i.e., the Fermi
energy is in the conduction band of surface state, and assume the
electric field is along the $x$ direction. To the lowest order of
the impurity density $n_i$ and stationary electric field $\bm
E=E\hat x$, the solution of the equation can be written as $\rho(\bm
k)=\rho^{(0)}(\bm k)+\rho^{(1)}(\bm k)+\rho^{(2)}(\bm
 k)$. Here $\rho^{(0)}(\bm k)={\rm diag}[n_{\rm F}(\varepsilon_{\bm k1}),n_{\rm F}(\varepsilon_{\bm
 k2})]$ [$n_{\rm F}(x)$ is the Fermi-Dirac function] is the equilibrium distribution function. $\rho^{(1)}(\bm k)$ and
$\rho^{(2)}(\bm k)$ are two distribution functions proportional to
the electric field. $\rho^{(1)}(\bm k)$ is the impurity-independent
distribution function and only the off-diagonal elements
$\rho^{(1)}_{12}(\bm k)=\rho^{(1)*}_{21}(\bm k)=\rho_{r}^{(1)}(\bm
k)
 +i\rho_{i}^{(1)}(\bm k)$ are nonzero, with
\begin{align}
\rho_{r}^{(1)}(\bm k)=&\frac{eE}{4k\epsilon_{\bm
k}}\sin2\beta_{\bm k}
\sin\theta_{\bm k}[n_{\rm F}(\varepsilon_{\bm k1})-n_{\rm F}(\varepsilon_{\bm k2})],\label{rho1r}\\
\rho_{i}^{(1)}(\bm k)=&\frac{eE}{16k\epsilon_{\bm
k}}\frac{\sin4\beta_{\bm k} (\cos4\theta_{\bm k}-5\cos2\theta_{\bm
k})}{\cos3\theta_{\bm k}} \nonumber\\&\times[n_{\rm
F}(\varepsilon_{\bm k1})-n_{\rm F}(\varepsilon_{\bm
k2})].\label{rho1i}
\end{align}
$\rho^{(2)}(\bm k)$ relies on the carrier-impurity scattering, and
its elements are determined by the following set of equations:
\begin{align}\label{}
eE\frac{\partial}{\partial k_x}n_{\rm F}(\varepsilon_{\bm k2})&=2\pi
n_i\sum_{\bm q} |V(\bm k-\bm q)|^2a_1(\bm k,\bm
q)\nonumber\\&\times\left[\rho_{22}^{(2)}
(\bm k)-\rho_{22}^{(2)}(\bm q)\right]\delta(\varepsilon_{\bm k2}-\varepsilon_{\bm q2}),\label{eq2}\\
2\epsilon_{\bm k}\rho_{i}^{(2)}(\bm k)&=\pi n_i\sum_{\bm q}|V(\bm
k-\bm q)|^2 a_2(\bm k,\bm
q)\nonumber\\&\times\left[\rho_{22}^{(2)}(\bm k)-\rho_{22}^{(2)}
(\bm q)\right]\delta(\varepsilon_{\bm k2}-\varepsilon_{\bm q2}),\label{eqr}\\
-2\epsilon_{\bm k}\rho_{r}^{(2)}(\bm k)&=\pi n_i\sum_{\bm q}|V(\bm
k-\bm q)|^2 a_3(\bm k,\bm
q)\nonumber\\&\times\left[\rho_{22}^{(2)}(\bm k)-\rho_{22}^{(2)}
(\bm q)\right]\delta(\varepsilon_{\bm k2}-\varepsilon_{\bm
q2}).\label{eqi}
\end{align}
$\rho_{r}^{(2)}(\bm k)$ and $\rho_{i}^{(2)}(\bm k)$ are the real and
imaginary parts of $\rho_{12}^{(2)}(\bm k)$. In these equations
\begin{align}\label{}
a_1(\bm k,\bm q)=&\frac{1}{2}\big[\sin2\beta_{\bm k}\sin2\beta_{\bm
q}
\cos(\theta_{\bm k}-\theta_{\bm q})\nonumber\\&+\cos2\beta_{\bm k}\cos2\beta_{\bm q}+1\big],\\
a_2(\bm k,\bm q)=&\frac{1}{2}\big[\cos2\beta_{\bm k}\sin2\beta_{\bm
q}
\cos(\theta_{\bm k}-\theta_{\bm q})\nonumber\\&-\sin2\beta_{\bm k}\cos2\beta_{\bm q}\big],\\
a_3(\bm k,\bm q)=&-\frac{1}{2}\sin2\beta_{\bm q}\sin(\theta_{\bm
k}-\theta_{\bm q}).
\end{align}
Note that the requirement that $\varepsilon_{\rm F}>0$, but the
Fermi energy is in the gap of bulk system is assumed, hence, the
diagonal element $\rho_{11}^{(2)} (\bm k)$ makes no contribution to
the transport equations. We find that when $\theta_{\bm
k}-\theta_{\bm q}=\pi$, $a_1(\bm k,\bm
q)=\tfrac{1}{2}[\cos(2\beta_{\bm k}-2\beta_{\bm q})+1]=0$. This
reveals the absence of backscattering, characteristic of TIs.

\subsection{conductivity and CISP}
In the eigenbasis of $\hat H_0$, the average velocity $\bm
v=\frac{1}{N}\sum_{\bm k}{\rm Tr}[\rho(\bm k)U_{\bm k}^\dag{\hat{\bm
v}} U_{\bm k}]$. Here two components of velocity operator in the
spin basis are written as
\begin{align}\label{}
\hat v_x=&\begin{pmatrix}
            3\lambda k^2\cos2\theta_{\bm k} & -iv_{\rm F} \\
            iv_{\rm F} & -3\lambda k^2\cos2\theta_{\bm k} \\
          \end{pmatrix},
\\
\hat v_y=&\begin{pmatrix}
            -3\lambda k^2\sin2\theta_{\bm k} & -v_{\rm F} \\
            -v_{\rm F} & 3\lambda k^2\sin2\theta_{\bm k} \\
          \end{pmatrix}.
\end{align}
It is seen that the diagonal elements of velocity operator are also
nonzero when we include the warping term. Therefore, the
longitudinal and transverse conductivities $\sigma_{xx}=-Nev_x/E$,
$\sigma_{xy}=-Nev_y/E$ can be expressed as
\begin{align}\label{}
\sigma_{xx}&=-\frac{e}{E}\sum_{\bm k}\Big[(v_{\rm F}\sin2\beta_{\bm
k}\cos\theta_{\bm k} \nonumber\\&+3\lambda k^2\cos2\beta_{\bm
k}\cos2\theta_{\bm k})\rho_{22}(\bm k)-2(v_{\rm F}\cos2\beta_{\bm
k}\cos\theta_{\bm k}
\nonumber\\
&-3\lambda k^2\sin2\beta_{\bm k}\cos2\theta_{\bm k})\rho_{r}(\bm k)
+2v_{\rm F}\sin\theta_{\bm k}\rho_{i}(\bm k)\Big],\label{sigma}\\
\sigma_{xy}&=-\frac{e}{E}\sum_{\bm k}\Big[(v_{\rm F}\sin2\beta_{\bm
k}\sin\theta_{\bm k} \nonumber\\&-3\lambda k^2\cos2\beta_{\bm
k}\sin2\theta_{\bm k})\rho_{22}(\bm k) -2(v_{\rm F}\cos2\beta_{\bm
k}\sin\theta_{\bm k}\nonumber\\& +3\lambda k^2\sin2\beta_{\bm
k}\sin2\theta_{\bm k})\rho_{r}(\bm k)-2v_{\rm F}\cos\theta_{\bm
k}\rho_{i}(\bm k)\Big].
\end{align}
The hexagonal warping term results in the complex forms of charge
conductivities. In the surface state of a TI, the carrier spin is
directly coupled to the momentum, in contrast to graphene. Hence, in
this system, an external in-plane electric field can lead to a
uniform spin polarization,\cite{Dimitrie} like spin-orbit-coupled
systems.
\cite{dyakonov1971cis,edelstein1990spc,wang2010spin,wang2010current}
This is the so called ``CISP". Three components of CISP $\bm
S=\sum_{\bm k}{\rm Tr}[\rho(\bm k)U_{\bm k}^\dag\frac{1}{2}\hat{\bm
\sigma} U_{\bm k}]$ are given by
\begin{align}\label{}
S_{x}=&\frac{1}{2}\sum_{\bm k}\Big[-\sin2\beta_{\bm
k}\sin\theta_{\bm k}\rho_{22}(\bm k)
+2\cos2\beta_{\bm k}\sin\theta_{\bm k}\rho_{r}(\bm k)\nonumber\\
&+2\cos\theta_{\bm k}\rho_{i}(\bm k)\Big],\\
S_{y}=&\frac{1}{2}\sum_{\bm k}\Big[\sin2\beta_{\bm k}\cos\theta_{\bm
k}\rho_{22}(\bm k)
-2\cos2\beta_{\bm k}\cos\theta_{\bm k}\rho_{r}(\bm k)\nonumber\\
&+2\sin\theta_{\bm k}\rho_{i}(\bm k)\Big],\\
S_{z}=&\frac{1}{2}\sum_{\bm k}\Big[\cos2\beta_{\bm k}\rho_{22}(\bm
k)+2\sin2\beta_{\bm k}\rho_{r}(\bm k)\Big].\label{sz}
\end{align}
It is noticeable that the general expressions
\eqref{sigma}-\eqref{sz} are applicable to any scattering potential.

According to the expressions \eqref{rho1r} and \eqref{rho1i}, it is
seen that $\rho_{r}^{(1)}(k_x,k_y)=(-1)^n\rho_{r}^{(1)}[(-1)^m
k_x,(-1)^n k_y]$ and $\rho_{i}^{(1)}(k_x,k_y)=\rho_{i}^{(1)}[(-1)^m
k_x,(-1)^n k_y]$ with $m,n=1,2$. We find that the
impurity-independent distribution makes no contribution to charge
conductivity and CISP. Furthermore, for normal nonmagnetic elastic
scattering, the potential satisfies the following
relation\cite{Yasuhiro7151}
\begin{equation}\label{Vrelation}
V(q,\theta_{\bm q})=V(q,\theta_{\bm q}-\pi)=V(q,\theta_{\bm q}+\pi).
\end{equation}
In connection with the kinetic equations \eqref{eq2}-\eqref{eqi},
one can directly arrive at the symmetrical relation:
$\rho_{22}^{(2)}(k_x,k_y)=(-1)^m\rho_{22}^{(2)}[(-1)^m k_x,(-1)^n
k_y]$, $\rho_{r}^{(2)}(k_x,k_y)=(-1)^n\rho_{r}^{(2)}[(-1)^m
k_x,(-1)^n k_y]$, and $\rho_{i}^{(2)}(k_x,k_y)=\rho_{i}^{(2)}[(-1)^m
k_x,(-1)^n k_y]$. Therefore, it is clear that $\sigma_{xy}=0$ and
$S_x=S_z=0$, and the off-diagonal elements of distribution function
have no effect on the charge conductivity and spin polarization.
Accordingly, the longitudinal conductivity $\sigma_{xx}$ and the
$y$-component of spin polarization can be rewritten as
\begin{align}\label{}
 \sigma_{xx}=&-\frac{e}{E}\sum_{\bm k}(v_{\rm F}\sin2\beta_{\bm k}\cos\theta_{\bm k}
 \nonumber\\&+3\lambda k^2\cos2\beta_{\bm k}\cos2\theta_{\bm k})\rho_{22}^{(2)}(\bm k),\label{sigxx}\\
S_{y}=&\frac{1}{2}\sum_{\bm k}\sin2\beta_{\bm k}\cos\theta_{\bm
k}\rho_{22}^{(2)}(\bm k).\label{sy}
\end{align}
For vanishing $\lambda$, the spin polarization linearly depends on the longitudinal
conductivity
\begin{equation}\label{}
\frac{\sigma_{xx}}{S_y}=-2\frac{ev_{\rm
F}}{E}.
\end{equation}
This relation is valid for any nonmagnetic elastic scattering. The
CISP can be observed using the Kerr rotation experiment.\cite{Kato}
Since CISP is the characteristic of surface state and there is no
spin polarization in bulk TIs, this relation may provide a simple
transport method to isolate the surface conductivity contribution in
Bi$_2$Se$_3$. One first measure the surface spin polarization, and
then the surface conductivity contribution can be obtained by this
relation. The remaining contribution of conductivity can be
considered to originate from bulk band. However, this method cannot
be applicable for Bi$_2$Te$_3$ due to its large warping effect.

The physical reason why only the $y$-component of CISP exists
is as follows. The CISP arises because an electric field
results in an average momentum $\langle\bm k\rangle=-e{\bm
E}\tau_{\rm tr}$ with $\tau_{\rm tr}$ being transport lifetime.
This implies from Hamiltonian \eqref{ham} that there is an
average spin-orbit field. This effective magnetic field leads
to this spin polarization. When the electric field is applied
along the $x$ direction, only the $y$- and $z$-components of
average effective magnetic field are nonzero. Further, the
$z$-component is a higher order term of electric field and
transport lifetime. Hence, in the limit of weak electric field
and weak scattering, only the $y$-component of spin
polarization is nonzero. We emphasize that this argument is
very general and valid for any scattering, including inelastic
phonon scattering.

\subsection{$\delta$-form short-range potential}
We first limit ourselves to a $\delta$-form short-range nonmagnetic
scattering $\widetilde V(\bm r)=\sum_{\{\bm R_\alpha\}}u\delta(\bm
r-\bm R_\alpha)$. This scattering arises from the surface roughness.
For this potential, the relation \eqref{Vrelation} is satisfied
explicitly. Hence, only the longitudinal conductivity and
$y$-component of CISP exist. To the second order of $\lambda$, the
diagonal element of matrix distribution function,
$\rho_{22}^{(2)}(\bm k)$, can be obtained analytically. At zero
temperature, it takes the form
\begin{align}\label{}
\rho_{22}^{(2)}(\bm k)=&-\frac{2eE}{n_iu^2}\bigg[\frac{2v_{\rm
F}^3}{\varepsilon_{\rm F}}\cos\theta_{\bm
k}+\lambda^2\frac{\varepsilon_{\rm F}^3}{4v_{\rm
F}^3}\big(18\cos\theta_{\bm k}\nonumber\\&+5\cos5\theta_{\bm
k}-\cos7\theta_{\bm k}\big)\bigg]\delta(\varepsilon_{\bm
k2}-\varepsilon_{\rm F}).
\end{align}
Substituting the resultant distribution function into Eqs.
\eqref{sigxx} and \eqref{sy}, the longitudinal conductivity
$\sigma_{xx}$ and spin polarization $S_y$ read
\begin{align}\label{}
\sigma_{xx}=&\frac{e^2}{\pi n_iu^2}\left[v_{\rm
F}^2+2\left(\frac{\varepsilon_{\rm F}}{v_{\rm
F}}\right)^4\lambda^2\right],\nonumber\\
=&\frac{e^2}{\pi n_iu^2}\left(v_{\rm F}^2+32\pi^2\lambda^2N^2\right),\\
S_y=&-\frac{eE}{4\pi n_iu^2}\left(2v_{\rm
F}+\frac{\varepsilon_{\rm F}^4}{v_{\rm
F}^5}\lambda^2\right),\nonumber\\
=&-\frac{eE}{2\pi v_{\rm F}n_iu^2}\left(v_{\rm
F}^2+8\pi^2\lambda^2N^2\right).
\end{align}
The hexagonal warping parameter $\lambda$ leads to quadratic
corrections of carrier density in the longitudinal conductivity
$\sigma_{xx}$ and CISP $S_y$. At the same time, the linear
relation between $\sigma_{xx}$ and $S_y$ is broken. We
emphasize here that the Hamiltonian for $\lambda=0$ used in
this paper is different from the one of Ref.
\onlinecite{Dimitrie} for $D=0$ by replacing $\hat
\sigma_y\rightarrow\hat\sigma_x$ and $\hat
\sigma_x\rightarrow-\hat\sigma_y$. Hence, for vanishing
$\lambda$, the above results are in agreement with previous
ones.\cite{Dimitrie} It is noticeable that the effective
Hamiltonian \eqref{ham} is obtained for low energy system and
the above two equations are valid in the density regime,
$10^{13}\,{\rm cm}^{-2}<N<10^{14}\,{\rm cm}^{-2}$, for
Bi$_2$Te$_3$. This is the precondition of the whole work.
Hence, all the equations are limited by this concealed
condition.

\subsection{screened Coulomb potential}
We now consider the screened Coulomb potential, where the screening
function is in the RPA $\epsilon_{\rm RPA}(\bm
q,\omega)=1-v_c(q)\Pi(\bm q;\omega)$ with
$v_c(q)=e^2/(2\epsilon_0\kappa q)$ being the two-dimensional Coulomb
interaction. The charged impurities scattering in the surface of TIs
can be modeled by this potential well. The corresponding
polarizability function takes the form:
\begin{align}\label{polafun}
\Pi(\bm q;\omega)=\sum_{\bm k,\mu,\mu'}&(\varphi_{\bm k+\bm
q\mu}^\dag\varphi_{\bm k\mu'})(\varphi_{\bm k\mu'}^\dag\varphi_{\bm
k+\bm q\mu})\nonumber\\&\times\frac{n_{\rm F}(\varepsilon_{\bm
k\mu'})-n_{\rm F}(\varepsilon_{\bm k+\bm
q\mu})}{\omega+\varepsilon_{\bm k\mu'}-\varepsilon_{\bm k+\bm
q\mu}+i\eta}.
\end{align}
The warping term complicates the calculation of the polarizability
function and we cannot get an analytical result even for static
case and vanishing temperature. The screened scattering potential is
related to the static dielectric function, and  is written as $V(\bm
q)=v_c(q)/\epsilon_{\rm RPA}(\bm q)$. Here the static dielectric
function $\epsilon_{\rm RPA}(\bm q)=1-v_c(q)\Pi(\bm q;0)$.

\subsubsection{static polarizability function}
The static polarizability $\Pi(\bm q;0)$ for $\varepsilon_{\rm F}$
to be in the conduction band of the surface of TI is given by
$\Pi(\bm q;0)=\Pi^+(\bm q;0)+\Pi^-(\bm q;0)$, where
\begin{align}\label{}
\Pi^+(\bm q;0)=&\sum_{\bm k}\bigg[a_1(\bm k,\bm k+\bm q)\frac{n_{\rm
F}(\varepsilon_{\bm k2})-n_{\rm F}(\varepsilon_{\bm k+\bm
q2})}{\varepsilon_{\bm k2}-\varepsilon_{\bm k+\bm
q2}}\nonumber\\
&+\bar{a}_1(\bm k,\bm k+\bm q)\frac{n_{\rm F}(\varepsilon_{\bm
k2})+n_{\rm F}(\varepsilon_{\bm k+\bm q2})}{\varepsilon_{\bm
k2}+\varepsilon_{\bm k+\bm q2}}\bigg],\label{pip}\\
\Pi^-(\bm q;0)=&\sum_{\bm k}\bigg[\bar{a}_1(\bm k,\bm k+\bm
q)\frac{n_{\rm F}(\varepsilon_{\bm k1})+n_{\rm F}(\varepsilon_{\bm
k+\bm q1})}{\varepsilon_{\bm k1}+\varepsilon_{\bm k+\bm q1}}\bigg].
\end{align}
Here $\bar{a}_1(\bm k,\bm k+\bm q)=1-a_1(\bm k,\bm k+\bm q)$. The
long-wavelength Thomas-Fermi (TF) screening is important for charged
impurity scattering. In the $q\rightarrow0$ limit, it is found that
$a_1(\bm k,\bm k+\bm q)\rightarrow1$. Hence,
$\Pi^-(q\rightarrow0,\theta_{\bm q};0)\rightarrow0$, and the
polarizability $\Pi(q\rightarrow0,\theta_{\bm q};0)$ is determined
by the first term of Eq. \eqref{pip}, and at zero temperature we
have
\begin{equation}\label{longwave}
\Pi(q\rightarrow0,\theta_{\bm q};0)=-\frac{\varepsilon_{\rm
F}}{2\pi^2}\int_0^{2\pi}d\theta_{\bm k}\frac{\cos(\theta_{\bm
k}-\theta_{\bm q})}{\Lambda(\theta_{\bm k},\theta_{\bm q})},
\end{equation}
where the angle-related function $\Lambda(\theta_{\bm k},\theta_{\bm
q})=2v_{\rm F}^2\cos(\theta_{\bm k}-\theta_{\bm q})+3\lambda^2k_{\rm
F}^4(\theta_{\bm k})[\cos(\theta_{\bm k}-\theta_{\bm
q})+\cos(5\theta_{\bm k}+\theta_{\bm q})]$. $k_{\rm F}(\theta_{\bm
k})$ is the Fermi momentum relying on the azimuthal angle, which is determined by
$\sqrt{v_{\rm F}^2k_{\rm F}^2(\theta_{\bm k})+\lambda^2 k_{\rm
F}^6(\theta_{\bm k})(\cos3\theta_{\bm k})^2}=\varepsilon_{\rm F}$.
For weak $\lambda$, it has the form:
\begin{equation}\label{kf}
k_{\rm F}(\theta_{\bm k})=\frac{\varepsilon_{\rm F}}{v_{\rm
F}}-\frac{1}{4}\frac{\lambda^2\varepsilon_{\rm F}^5}{v_{\rm
F}^7}(1+\cos6\theta_{\bm k}).
\end{equation}
In $\lambda\rightarrow0$ limit, the Fermi momentum tends to the
previous result.\cite{Dimitrie} According to
$N=\frac{1}{8\pi^2}\int_0^{2\pi}d\theta_{\bm k}k_{\rm
F}^2(\theta_{\bm k})$, the relation between the carrier density and
Fermi energy is given by
\begin{equation}\label{}
N=\frac{1}{4\pi}\left[\left(\frac{\varepsilon_{\rm F}}{v_{\rm
F}}\right)^2-\frac{1}{2}\frac{\varepsilon_{\rm F}^6}{v_{\rm
F}^8}\lambda^2\right].
\end{equation}
It should be noted that for our case only the
magnitude of momentum $\bm q$ tends to zero in the long-wavelength limit. Therefore, the
resultant polarizability $\Pi(q\rightarrow0,\theta_{\bm q};0)$ may
still rely on the azimuthal angle of $\bm q$, which is completely
different from the 2D Lindhard function\cite{Stern1967} and the
corresponding polarizability function of
graphene.\cite{wunsch2006dynamical,Hwang2007} For weak $\lambda$,
the integral \eqref{longwave} can be calculated, analytically, and
the TF dielectric function reduces to
\begin{equation}\label{}
\epsilon_{\rm TF}(\bm q)=1+\frac{k_{\rm TF}(\theta_{\bm q})}{q},
\end{equation}
with the angle-dependent TF wave vector $k_{\rm
TF}(\theta_{\bm q})$
\begin{equation}\label{}
k_{\rm TF}(\theta_{\bm q})=\frac{e^2\varepsilon_{\rm
F}}{8\pi\epsilon_0\kappa v_{\rm F}^8}\Big[2v_{\rm
F}^6-3\lambda^2\varepsilon_{\rm F}^4(1+\cos6\theta_{\bm q})\Big].
\end{equation}
This dielectric function $\epsilon_{\rm TF}(\bm q)$ tends to the
previous result\cite{Hwang2007,Dimitrie} when $\lambda\rightarrow0$.
We emphasize again that the angle dependence of dielectric function originates
 from the cubic term in the Hamiltonian. The $\lambda$-related term
becomes important when $v_{\rm F}^6=\lambda^2\varepsilon_{\rm F}^4$,
corresponding to $\varepsilon_{\rm F}=0.26\,{\rm eV}$ in
Bi$_2$Te$_3$, a density of $10^{13}\,{\rm cm}^{-2}$, which is a
realistic density in Bi$_2$Te$_3$ sample.\cite{Chen10072009}

\subsubsection{numerical results}\label{num}

\begin{figure}
\begin{center}
  \includegraphics[width=0.45\textwidth]{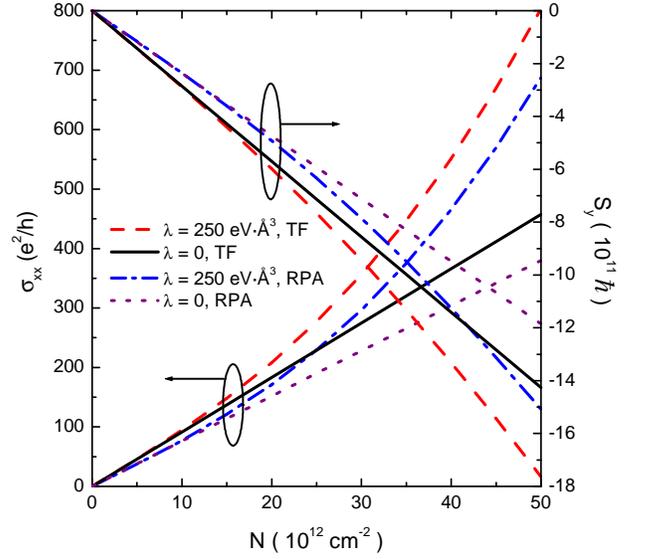}
\end{center}
\caption{(Color online) The longitudinal conductivity and CISP scattered by
screened charged impurities as a function of the surface carrier
density for both TF screening and RPA screening.}\label{res}
\end{figure}

It is seen that the polarizability function \eqref{polafun}
satisfies $\Pi(q,\theta_{\bm q};\omega)=\Pi(q,\theta_{\bm
q}-\pi;\omega)=\Pi(q,\theta_{\bm q}+\pi;\omega)$. Consequently, the
angle relation of scattering potential \eqref{Vrelation} still holds
for this screened Coulomb potential. Hence, when the carriers are
scattered by the screened Coulomb form, only the longitudinal
conductivity and $y$-component of CISP are nonzero. Now we
numerically calculate the longitudinal conductivity and spin
polarization for both TF screening and RPA screening. The following
parameters in Bi$_2$Te$_3$ are used in the
calculation:\cite{Fu266801,Dimitrie} Fermi velocity $v_{\rm
F}=2.55\,{\rm eV\cdot\AA}$, warping parameter $\lambda=250\,{\rm
eV\cdot\AA^3}$, impurity density $n_i=10^{13}\,{\rm cm}^{-2}$, the
static dielectric constant $\kappa=200$, and dc electric field
$E=10\,{\rm V/m}$. The results are plotted in Fig. \ref{res}. The
corresponding conductivity and CISP for $\lambda=0$ are also plotted
for comparison. For vanishing $\lambda$, the conductivity and spin
polarization linearly rely on the surface carrier density $N$, which
agrees with the previous theoretical calculation.\cite{Dimitrie}
With increasing the density $N$, the hexagonal warping effect
becomes important for both TF and RPA screened Coulomb potential,
leading to nonlinear characters of conductivity and CISP. The
magnitudes of conductivity and spin polarization in the presence of
warping effect are larger than the ones in the absence of warping.
Furthermore, we verify that, $\sigma_{xx}=c_1N+c_2N^2$,
$S_y=c_3N+c_4N^2$, in contrast to the ones of short-range
scattering. We note that the TF screening is the long-wavelength
limit of RPA dielectric function. At the same time, for finite
magnitude of momentum, the RPA screening is weaker than the TF one.
Hence, the conductivity for TF screened Coulomb potential is larger
than the RPA one at the same $\lambda$.

Here we have assumed that the impurities are located right on the
surface. However, the charged impurities in the bulk of TIs may also
contribute to the surface transport. These remote impurities will
enhance the magnitude of conductivity and spin polarization. One can
deduce that the warping will lead to a large increase of the
magnitude of surface transport quantities even in the presence of
remote impurities.

Note that the classical surface conductivity of Bi$_2$Se$_3$
was observed by using a gate electrode.\cite{Kim} It showed a
linear carrier density dependence when the density is smaller
than the carrier density above which the bulk conduction band
is populated. In Bi$_2$Se$_3$, the hexagonal warping is small
enough to be omitted completely. Hence, this experiment
verified the previous theoretical prediction
well.\cite{Dimitrie} However, for the surface states of TI
where the warping term cannot be neglected, such as
Bi$_2$Te$_3$, the linear dependence will be broken and a
quadratic relation also appears. Therefore, our prediction
suggests that the surface transport in Bi$_2$Te$_3$ should be
different from the one of Bi$_2$Se$_3$ and it should be very
careful when one analyzes the surface transport data of
Bi$_2$Te$_3$.

\section{classical transport in the presence of magnetic scattering}
Let us now address the magnetic scattering case in the classical
diffusion regime, where the scattering potential
reads\cite{Liu156603}
\begin{align}\label{}
\widetilde V(\bm r)=&\sum_{\{\bm R_\alpha\}}\{J_\parallel [s_x(\bm
r)\tilde{S}_x(\bm R_\alpha)+s_y(\bm r)\tilde{S}_y(\bm
R_\alpha)]\nonumber\\&+J_zs_z(\bm r)\tilde{S}_z(\bm R_\alpha)\}\delta(\bm r-\bm
R_\alpha).
\end{align}
Here $\bm s=\frac{1}{2}\bm \sigma$ is the spin vector of electron,
and ${\tilde{\bm S}}$ is the impurity spin, and $J_\parallel$, $J_z$
are the coupling parameters.

For simplicity, we assume the classical magnetic impurities and
their spins polarized in the $z$ direction. This kind of potential
conserves the $z$-component of the carrier spin. It is known that
magnetic doping will open a gap in the helical Dirac
cone.\cite{chensci} From the mean field approximation, the gap has
the form:\cite{culcer2010anomalous} $\Delta=2n_iJ_z\tilde{S}$.
Hence, for high mobility sample, the density of magnetic impurities
is small enough, and then the gap opened by the magnetic doping is
considered to be very small. When $\Delta\ll\varepsilon_{\rm F}$,
the effect of gap on the energy spectrum, group velocity, etc. could
be neglected safely. Therefore, we can only consider the scattering
effect of magnetic impurities. Applying the similar procedure as the
nonmagnetic scattering situation, the analogous kinetic equations
can be derived only by replacing $V(\bm k-\bm q)$, $a_1(\bm k,\bm
q)$, $a_2(\bm k,\bm q)$, and $a_3(\bm k,\bm q)$ in Eqs.
\eqref{eq2}-\eqref{eqi} with $u_{\rm M}$, $a_1^{\rm M}(\bm k,\bm
q)$, $a_2^{\rm M}(\bm k,\bm q)$, and $a_3^{\rm M}(\bm k,\bm q)$.
Here $u_{\rm M}=J_z\tilde{S}/2$ and
\begin{align}\label{}
a_1^{\rm M}(\bm k,\bm q)=&-\frac{1}{2}\big[\sin2\beta_{\bm
k}\sin2\beta_{\bm q}
\cos(\theta_{\bm k}-\theta_{\bm q})\nonumber\\&-\cos2\beta_{\bm k}\cos2\beta_{\bm q}-1\big],\\
a_2^{\rm M}(\bm k,\bm q)=&-\frac{1}{2}\big[\cos2\beta_{\bm
k}\sin2\beta_{\bm q}
\cos(\theta_{\bm k}-\theta_{\bm q})\nonumber\\&+\sin2\beta_{\bm k}\cos2\beta_{\bm q}\big],\\
a_3^{\rm M}(\bm k,\bm q)=&\frac{1}{2}\sin2\beta_{\bm
q}\sin(\theta_{\bm k}-\theta_{\bm q}).
\end{align}
Taking into account the symmetrical property of distribution
function, it is also verified that only the longitudinal
conductivity and $y$-component of CISP are nonzero for this magnetic
scattering. Eventually, the expressions of them are given by Eqs.
\eqref{sigxx} and \eqref{sy}.

We first assume that the warping parameter is weak. Thus the kinetic
equation can be solved analytically and the diagonal element of the
impurity-related distribution is
\begin{align}\label{}
\rho_{22}^{(2)}(\bm k)=&-\frac{2eE}{3n_iu_{\rm
M}^2}\Bigg[\frac{2v_{\rm F}^3}{\varepsilon_{\rm F}}\cos\theta_{\bm
k}+\lambda^2\frac{\varepsilon_{\rm F}^3}{2v_{\rm
F}^3}\bigg(\frac{35}{3}\cos\theta_{\bm
k}\nonumber\\&+\frac{17}{2}\cos5\theta_{\bm
k}-\frac{1}{2}\cos7\theta_{\bm k}\bigg)\Bigg]\delta(\varepsilon_{\bm
k2}-\varepsilon_{\rm F}).
\end{align}
Hence, the charge conductivity and CISP are written as
\begin{align}
\sigma_{xx}=&\frac{e^2}{9\pi n_iu^2_{\rm M}}\left[3v_{\rm
F}^2+8\left(\frac{\varepsilon_{\rm F}}{v_{\rm
F}}\right)^4\lambda^2\right],\nonumber\\
=&\frac{e^2}{9\pi n_iu^2_{\rm M}}\left(3v_{\rm F}^2+128\pi^2\lambda^2N^2\right),\label{weaksig}\\
S_y=&-\frac{eE}{36\pi n_iu^2_{\rm M}}\left(6v_{\rm
F}+7\frac{\varepsilon_{\rm F}^4}{v_{\rm
F}^5}\lambda^2\right),\nonumber\\
=&-\frac{eE}{18\pi v_{\rm F}n_iu_{\rm M}^2}\left(3v_{\rm
F}^2+56\pi^2\lambda^2N^2\right).\label{weaksy}
\end{align}
Compared with the short-range nonmagnetic scattering, similar
density-dependent behaviors have also been seen for this magnetic
one. However, the concrete coefficients are distinct, completely.

\begin{figure}
\begin{center}
  \includegraphics[width=0.45\textwidth]{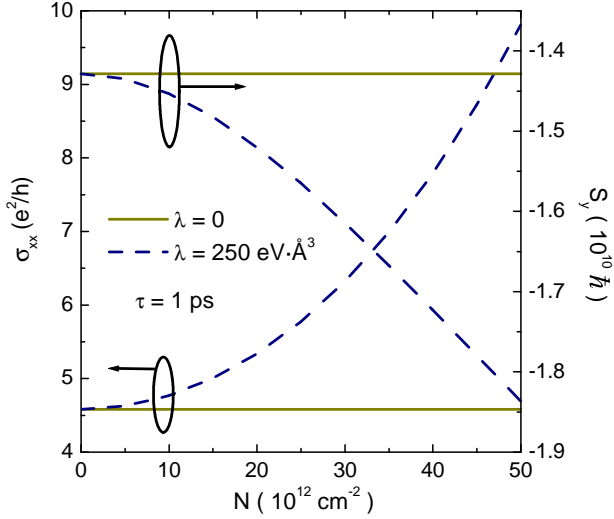}
\end{center}
\caption{(Color online) The longitudinal conductivity and $y$-component of CISP scattered by
magnetic impurities as a function of the surface carrier density.
The other parameters are the same as the ones in Sec.
\ref{num}.}\label{mag}
\end{figure}

To go beyond the weak warping case, we now numerically solve the
kinetic equations. Setting the relaxation time $\tau={2v_{\rm
F}}/({n_iu_{\rm M}^2\sqrt{4\pi N_0}})=1\,{\rm ps}$ with $N_0=10^{12}\,{\rm
cm}^{-2}$, the obtained longitudinal conductivity and spin
polarization are plotted in Fig. \ref{mag}. For comparison, the
conductivity and CISP without warping effect are also calculated. It
is seen that for fixed relaxation time the warping term also has
important role on the surface transport of a three-dimensional TI.
The magnitude of longitudinal conductivity and spin polarization
increase drastically with increasing the surface density.
Note that the above analytical results Eqs. \eqref{weaksig} and
\eqref{weaksy} are valid for weak warping. That is a density $N\ll
v_{\rm F}/(5\pi\lambda)\approx5\times 10^{12}\,{\rm cm}^{-2}$. For
example, if we use the approximation result \eqref{weaksig} to
estimate the conductivity, the resultant
$\sigma_{xx}\approx171.3\,{\rm e^2/h}$ when $N=30\times
10^{12}\,{\rm cm}^{-2}$. This value is much larger than the
numerical one. At the same time, it can be confirmed from the
 numerical calculation that the additional terms $\varpropto N$
also contribute to conductivity and spin polarization for
nonvanishing warping.

\section{quantum correction}
\begin{figure}
\begin{center}
  \includegraphics[width=0.45\textwidth]{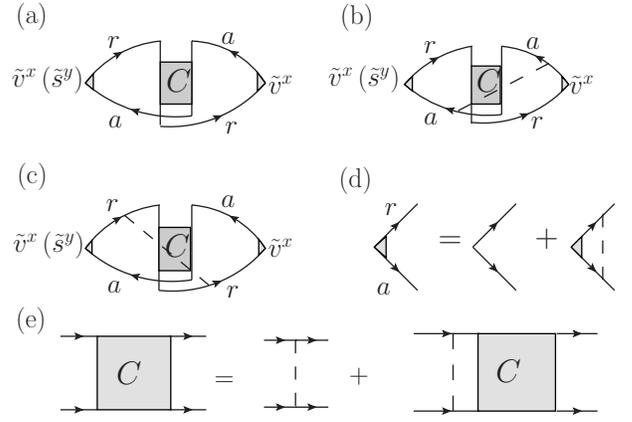}
\end{center}
\caption{The diagrams for the quantum corrections to surface
conductivity and spin polarization of TI. (a) Bare and [(b)-(c)] two
dressed Hikami boxes. (d) Equation for the vertex correction to
velocity and average spin in the ladder approximation. (e)
Bethe-Salpeter equation for the Cooperon. The arrowed solid and
dashed lines represent the retarded and advanced Green's functions,
and scattering potential, respectively.}\label{feyn}
\end{figure}

We now focus on the effect of weak warping on the quantum
corrections to conductivity and spin polarization. For this
surface state, the Berry phase is calculated as
\begin{align}\label{}
\gamma=&-i\int_0^{2\pi}d\theta_{\bm k}\langle \varphi_{\bm
k2}|\frac{\partial}{\partial \theta_{\bm k}}| \varphi_{\bm
k2}\rangle\nonumber\\=&\frac{1}{2}\int_0^{2\pi}d\theta_{\bm
k}\left[1+\frac{\lambda k_{\rm F}^3(\theta_{\bm
k})}{\varepsilon_{\rm F}}\cos3\theta_{\bm k}\right].
\end{align}
In connection with Eq. \eqref{kf}, the Berry phase equals $\pi$.
Hence, the weak anti-localization is expected for the surface state
even in the presence of warping effect.

Using equilibrium Green's function, the quantum corrections are
described by the diagrams in Fig. \ref{feyn}. Firstly, we
consider the short-range nonmagnetic scattering. Note that in
the previous work the authors replaced the energy spectrum by
its angle average one to investigate the quantum correction. In
our situation the angle dependence of warping is taken into
account, hence, our treatment is beyond this approximation. We
also assume that the Fermi energy is in the gap of bulk band,
and crosses the upper band of surface state. Under Born
approximation, the impurity-averaged equilibrium retarded and
advanced Green's functions are given by
\begin{equation}\label{}
{\hat G}^{r/a}_{\bm k}(\epsilon)=\frac{1}{\epsilon-\epsilon_{\bm k}\pm
i/2\tau_e},
\end{equation}
with the relaxation time
$\tfrac{1}{\tau_e}=n_iu^2\tfrac{\varepsilon_{\rm F}}{4v_{\rm
F}^8}(2v_{\rm F}^6-3\varepsilon_{\rm F}^4\lambda^2)$. We add a hat
on the equilibrium Green's functions to distinguish them from the
nonequilibrium Green's functions. Notice that we have used a matrix
distribution function to discuss the classical transport. However,
in the absence of interband transition process and for usual elastic
scattering, the matrix distribution reduces to a scalar one [see
Eqs. \eqref{eq2}, \eqref{sigxx} and \eqref{sy}]. Hence, the kinetic
equation approach is in principle equivalent to the one-band
equilibrium Green's functions approach. The kinetic equation
approach can easily deal with momentum-dependent scattering in
classical transport, but it is difficult to discuss the quantum
correction. Therefore, we treat the weak anti-localization in the
diagrammatic approach by using equilibrium Green's functions. Below,
the word ``equilibrium" will be omitted for brevity.

In the calculation, the vertex corrections to the bare velocity and
average spin Fig. \ref{feyn} (d) should be taken into account, which
are written as
\begin{align}\label{}
\tilde{v}^x_{\bm k}&=2v_{\rm F}\cos\theta_{\bm k}+\frac{\lambda^2
k^4}{2v_{\rm F}}\left(3\cos\theta_{\bm
k}+\frac{5}{2}\cos5\theta_{\bm k}-\frac{1}{2}\cos7\theta_{\bm
k}\right),\\
\tilde{s}^y_{\bm k}&=\cos\theta_{\bm k}-\frac{\lambda^2 k^4}{4v_{\rm
F}^2}\left(3\cos\theta_{\bm k}+\frac{1}{2}\cos5\theta_{\bm
k}+\frac{1}{2}\cos7\theta_{\bm k}\right).
\end{align}
Different from the topological surface states in the absence of
hexagonal warping, the velocity and spin vertex become anisotropic.
This anisotropy is very important and will lead to the density
dependence of quantum correction. Note that in the absence of
warping the group velocity $v^x_{\bm k}={\partial \epsilon_{\bm
k}}/{\partial k_x}|_{\lambda=0}=v_{\rm F}\cos\theta_{\bm k}$ and
average spin $s^y_{\bm k}=\tfrac{1}{2}\langle\varphi_{\bm k2}
|\hat{\sigma}_y|\varphi_{\bm
k2}\rangle|_{\lambda=0}=\tfrac{1}{2}\cos\theta_{\bm k}$. Therefore,
for vanishing $\lambda$, the velocity vertex reduces
to\cite{McCann2006} $\tilde{v}^x_{\bm k}=2v^x_{\bm k}$.

In addition to the bare Hikami box Fig. \ref{feyn} (a), two dresses
Hikami boxes Fig. \ref{feyn} (b)-(c) are also needed in the
calculation of quantum corrections and the total corrections of
charge conductivity is given by
\begin{equation}\label{}
\delta\sigma_{xx}=\delta\sigma_{xx}^{(1)}+\delta\sigma_{xx}^{(2)}+\delta\sigma_{xx}^{(3)},
\end{equation}
where the quantum correction due to bare Hikami box
\begin{equation}\label{}
 \delta\sigma_{xx}^{(1)}=\frac{e^2}{2\pi}\sum_{\bm k,\bm q}\tilde{v}^x_{\bm k}
 \hat G_{\bm k}^r \hat G_{\bm q-\bm k}^r \tilde{v}^x_{\bm q-\bm k} \hat G_{\bm q-\bm k}^a\hat G_{\bm k}^a C(\bm q),
\end{equation}
and the quantum corrections due to two dressed Hikami boxes
\begin{align}\label{}
  \delta\sigma_{xx}^{(2)}=&\frac{e^2}{2\pi}\sum_{\bm k,\bm q,\bm k'}\tilde{v}^x_{\bm k}
 \hat G_{\bm k}^r \hat G_{\bm k'}^r\langle|\xi_{\bm k\bm k'}|^2\rangle_{\rm imp} \hat G_{\bm q-\bm k}^r \hat G_{\bm q-\bm k'}^r
 \nonumber\\&\times\tilde{v}^x_{\bm q-\bm k'} \hat G_{\bm q-\bm k'}^a\hat G_{\bm k}^a C(\bm q),\\
  \delta\sigma_{xx}^{(3)}=&\frac{e^2}{2\pi}\sum_{\bm k,\bm q,\bm k'}\tilde{v}^x_{\bm k}
 \hat G_{\bm k}^r \hat G_{\bm q-\bm k'}^r\langle|\xi_{\bm k\bm k'}|^2\rangle_{\rm imp} \hat G_{\bm q-\bm k'}^a \hat G_{\bm q-\bm
 k}^a
 \nonumber\\&\times\tilde{v}^x_{\bm q-\bm k'} \hat G_{\bm k'}^a\hat G_{\bm k}^a C(\bm q).
\end{align}
Here $\xi_{\bm k\bm k'}=\int d\bm r e^{i(\bm k'-\bm k)\cdot\bm
r}\langle\varphi_{\bm k2}|\widetilde V(\bm r)|\varphi_{\bm
k'2}\rangle$ is the scattering amplitude between two eigenstates.
$\langle\cdots\rangle_{\rm imp}$ means average over all possible
configurations of random impurity. Consequently, for $\delta$-form
impurity scattering, $\langle|\xi_{\bm k\bm k'}|^2\rangle_{\rm
imp}=n_iu^2\langle\varphi_{\bm k2}|\varphi_{\bm
k'2}\rangle\langle\varphi_{\bm k'2}|\varphi_{\bm k2}\rangle$. Since
the Cooperon $C(\bm q)$ diverges as $q\rightarrow0$, hence the most
divergent terms could be obtained by setting $\bm q=0$ for velocity
vertex, retarded and advanced Green's functions in the above
expressions. Performing momentum $\bm k$ and $\bm k'$ integrals, the
total conductivity correction is written as
\begin{equation}\label{}
\delta\sigma_{xx}=-\frac{e^2}{\pi}\varepsilon_{\rm
F}\tau_e^3\left(1+\frac{\lambda^2\varepsilon_{\rm F}^4}{2v_{\rm
F}^6}\right)\sum_{\bm q}C(\bm q).
\end{equation}
For vanishing $\lambda$, this result is in accordance with the one
in Ref. \onlinecite{lu2011competing}. Similarly, the total spin
polarization correction is given by
\begin{equation}\label{}
\delta S_y=\frac{eE}{2\pi}\frac{\varepsilon_{\rm F}\tau_e^3}{v_{\rm
F}}\left(1-\frac{\lambda^2\varepsilon_{\rm F}^4}{v_{\rm
F}^6}\right)\sum_{\bm q}C(\bm q).
\end{equation}
The Cooperon satisfies the Bethe-Salpeter equation Fig. \ref{feyn}
(e)
\begin{align}\label{BSeq}
C_{\bm k_1\bm k_2}=&C_{\bm k_1\bm k_2}^{(0)}+\sum_{\bm k'}C_{\bm
k_1\bm k'}^{(0)}G_{\bm k'}^r G_{\bm q-\bm k'}^a C_{\bm k'\bm
k_2}\nonumber\\
=&C_{\bm k_1\bm k_2}^{(0)}+\int \frac{dk'd\theta_{\bm
k'}}{(2\pi)^2}F(\bm k_1,\bm k',\bm q)C_{\bm k'\bm k_2},
\end{align}
where $\bm k_1+\bm k_2=\bm q$, $F(\bm k_1,\bm k',\bm q)=k'C_{\bm
k_1\bm k'}^{(0)}\hat G_{\bm k'}^r\hat G_{\bm q-\bm k'}^a$, and by expanding up
to $\lambda^2$, the bare vertex $C_{\bm k_1\bm k_2}^{(0)}$ for small
$q$ is written as
\begin{equation}\label{}
C_{\bm k_1\bm k_2}^{(0)}=\Upsilon^{(0)}_{\bm k_1\bm
k_2}+\lambda^2\Upsilon^{(2)}_{\bm k_1\bm k_2},
\end{equation}
with
\begin{align}\label{}
\Upsilon^{(0)}_{\bm k_1\bm k_2}=\frac{v_{\rm F}^2}{2\varepsilon_{\rm
F}\tau_{e}}\left[1+2e^{i(\theta_{\bm k_1} -\theta_{\bm
k_2})}+e^{2i(\theta_{\bm k_1}-\theta_{\bm k_2})}\right].
\end{align}
The expression of $\Upsilon^{(2)}_{\bm k_1\bm k_2}$ is long and we
do not write it here. For weak warping, the solution of Eq.
\eqref{BSeq} is found as $C_{\bm k_1\bm k_2}=\Lambda_{\bm k_1\bm
k_2}^{(0)}+\lambda^2\Lambda_{\bm k_1\bm k_2}^{(2)}$. $\Lambda_{\bm
k_1\bm k_2}^{(0)}$ and $\Lambda_{\bm k_1\bm k_2}^{(2)}$ are
independent of $\lambda$. With the help of the expansion $F(\bm
k_1,\bm k',\bm q)=F^{(0)}(\bm k_1,\bm k',\bm q)+\lambda^2F^{(2)}(\bm
k_1,\bm k',\bm q)$, $\Lambda_{\bm k_1\bm k_2}^{(0)}$ and
$\Lambda_{\bm k_1\bm k_2}^{(2)}$ are determined by the following
equations:
\begin{align}\label{}
\Lambda_{\bm k_1\bm k_2}^{(0)}=&\Upsilon^{(0)}_{\bm k_1\bm k_2}+\int
\frac{dk'd\theta_{\bm
k'}}{(2\pi)^2}F^{(0)}(\bm k_1,\bm k',\bm q)\Lambda_{\bm k_1\bm k_2}^{(0)}, \label{eqz}\\
\Lambda_{\bm k_1\bm k_2}^{(2)}=&\Upsilon^{(2)}_{\bm k_1\bm k_2}+\int
\frac{dk'd\theta_{\bm k'}}{(2\pi)^2}F^{(2)}(\bm k_1,\bm k',\bm
q)\Lambda_{\bm k'\bm k_2}^{(0)}\nonumber\\&+\int
\frac{dk'd\theta_{\bm k'}}{(2\pi)^2}F^{(0)}(\bm k_1,\bm k',\bm
q)\Lambda_{\bm k'\bm k_2}^{(2)}.\label{eqh}
\end{align}
$\Lambda_{\bm k_1\bm k_2}^{(0)}$ can be acquired from Eq. \eqref{eqz}
and then we can obtain $\Lambda_{\bm k_1\bm k_2}^{(2)}$ From Eq.
\eqref{eqh}. It is found that $F^{(0)}(\bm k_1,\bm k',\bm q)$ relies
on $\theta_{\bm k_1}$ through $\cos\theta_{\bm k_1}$,
$\sin\theta_{\bm k_1}$, $\cos2\theta_{\bm k_1}$, and
$\sin2\theta_{\bm k_1}$. Hence, $\Lambda_{\bm k_1\bm k_2}^{(0)}$ and
$\Lambda_{\bm k_1\bm k_2}^{(2)}$ have the forms:
\begin{align}
  \Lambda_{\bm k_1\bm k_2}^{(0)} =& \Upsilon^{(0)}_{\bm k_1\bm k_2}+
  \mathcal A_0+\mathcal A_1\cos\theta_{\bm k_1}+\mathcal B_1\sin\theta_{\bm k_1}
  \nonumber\\&+\mathcal A_2\cos2\theta_{\bm k_1}+\mathcal B_2\sin2\theta_{\bm k_1},\\
  \Lambda_{\bm k_1\bm k_2}^{(2)} =& \Upsilon^{(2)}_{\bm k_1\bm k_2}+\int
\frac{dk'd\theta_{\bm k'}}{(2\pi)^2}F^{(2)}(\bm k_1,\bm k',\bm
q)\Lambda_{\bm k'\bm k_2}^{(0)}\nonumber\\&+\mathcal C_0+\mathcal
C_1\cos\theta_{\bm k_1}+\mathcal D_1\sin\theta_{\bm k_1}
  \nonumber\\&+\mathcal C_2\cos2\theta_{\bm k_1}+\mathcal D_2\sin2\theta_{\bm k_1}.
\end{align}
The above coefficients $\mathcal A_i$, $\mathcal B_j$, $\mathcal
C_i$, and $\mathcal D_j$ ($i=0,1,2$, $j=1,2$) are independent of
$\theta_{\bm k_1}$ and can be determined by Eqs. \eqref{eqz} and
\eqref{eqh}. The derivation is tedious but direct. By setting $\bm
k_1=\bm k$, $\bm k_2=\bm q-\bm k$ and $\theta_{\bm k_1}-\theta_{\bm
k_2}=\pi$ for small $q$ and collecting the most divergent terms,
finally, the Cooperon is obtained as
\begin{equation}\label{}
C(\bm q)=-\frac{1}{\varepsilon_{\rm
F}\tau_e^3q^2}-\frac{5}{4}\frac{\lambda^2\varepsilon_{\rm
F}^3}{v_{\rm F}^6\tau_{e}^3q^2}.
\end{equation}
By performing the integration over $q$ between $1/l_{\phi}$ and
$1/l$, the logarithmic correction to the conductivity and spin
polarization are found as
\begin{align}\label{}
\delta\sigma_{xx}=&\frac{e^2}{4\pi^2}\left(1+\frac{3}{4}\frac{\lambda^2
\varepsilon_{\rm F}^4}{v_{\rm F}^6}\right)\ln\frac{\tau_\phi}{\tau_e},\\
\delta S_y=&-\frac{eE}{8\pi^2v_{\rm
F}}\left(1+\frac{1}{4}\frac{\lambda^2 \varepsilon_{\rm F}^4}{v_{\rm
F}^6}\right)\ln\frac{\tau_\phi}{\tau_e}.
\end{align}
Here we have used the relations $l=\sqrt{D\tau_e}$ and
$l_\phi=\sqrt{D\tau_\phi}$ with $D$ the diffusion constant. It
is useful to rewrite the conductivity correction as
$\delta\sigma_{xx}=-\alpha\tfrac{e^2}{2\pi^2}\ln\frac{\tau_\phi}{\tau_e}$.
Therefore, the $\alpha$ has the form:
\begin{align}\label{}
\alpha=&-\frac{1}{2}\left(1+\frac{3}{4}\frac{\lambda^2
\varepsilon_{\rm F}^4}{v_{\rm F}^6}\right),\nonumber\\
=&-\frac{1}{2}\left(1+\frac{12\pi^2\lambda^2}{v_{\rm
F}^2}N^2\right).
\end{align}
The hexagonal warping makes the prefactor $\alpha$ quadratically
depend on the carrier density and is always smaller than
$-\tfrac{1}{2}$, in vivid contrast with the angle-average
approximation.\cite{tkachov2011weak} For vanishing warping, this
factor becomes $-\tfrac{1}{2}$, in agreement with the theoretical
work.\cite{lu2011competing} We note that the above formula is
fulfilled for weak warping. On the other hand, one can estimate the
$\alpha$ for Bi$_2$Te$_3$ at low surface density $N\ll10^{13}\,{\rm
cm}^{-2}$. For instance, $\alpha\approx-0.506$ when $N=10^{12}\,{\rm
cm}^{-2}$. However, the value of the $\alpha$ obtained from fit in
recent experiment\cite{HongTao2011} is $-0.39$, which is larger than
$-0.5$, conflicting with our formula. This may be due to the
inevitable bulk states contribution in three-dimensional
TIs.\cite{lu2011weak} The experimental observation of $\alpha$ is a
collective result of the surface bands and bulk bands. The bulk
channels may result in a weak localization term, which could reduce
or even compensate the weak anti-localization arising from the
surface states. Therefore, a larger value of $\alpha$ is obtained
experimentally. In contrast to the surface bands, the bulk subbands
of TIs have a quadratic term and large band gaps. As a result, a
different density-dependent behavior of quantum correction is
expected for the bulk channels. A quantitative measurement of the
carrier-density-dependent surface conductivity correction could be
helpful for distinguishing the surface contribution from the bulk
one.

In the presence of magnetic scattering, the divergence of $C(\bm q)$
when $q\rightarrow0$ vanishes, which is analogous to the case
without warping.\cite{lu2011competing} Therefore the logarithmic
correction disappears, and it could be deduced that the magnetic
scattering suppresses the weak anti-localization effect in the
presence of both magnetic and nonmagnetic scattering. This is in
accordance with experimental observation.\cite{HongTao2011}

\section{conclusion}
In summary, we have investigated the surface transport of a
three-dimensional TI both in classical and quantum diffusive
regimes. In this study, we include the role of the hexagonal warping
correction of Fermi surface. It is found that the hexagonal warping
has drastic effects on the surface conductivity and CISP of a
three-dimensional TI for both nonmagnetic and magnetic elastic
scattering. For surface state with large warping, such as
Bi$_2$Te$_3$, an additional quadratic carrier density dependence is
found in both two regimes. Because the carrier density could be
controlled by the gate voltage, hence, we hope that our predictions
will soon be verified experimentally.

\begin{acknowledgments}
This work was supported by the National Science Foundation of China
(Grants No. 11104002 and No. 60876064).
\end{acknowledgments}

\end{document}